\documentclass[authoryear,review]{elsarticle}
\journal{Journal of Systems and Software}

\newcommand{\CE}{Continuous Experimentation}
\newcommand{\cps}{cyber-physical systems}

\newcommand{\RQone}{In the context of \cps, what is the state-of-the-art of \CE?}
\newcommand{\RQtwo}{In the context of \cps\ and more specifically the automotive industry, what feedback do the practitioners provide about the \CE\ practice?}

\newcommand{\RGoal}{This paper aims at providing an overview of the engagement on the \CE\ practice in the context of \cps}

\usepackage{lineno}
\modulolinenumbers[5]

\usepackage{hyperref}
\usepackage[capitalise]{cleveref}
\usepackage{enumitem}

\usepackage{array}

\usepackage{ifthen,color,amssymb}

\usepackage{pdflscape}
\usepackage{tabularx}
\usepackage{multirow}
\usepackage{afterpage}

\usepackage{tikz}
\usepackage{float}

\newcommand*\justify{%
  \fontdimen2\font=0.4em
  \fontdimen3\font=0.2em
  \fontdimen4\font=0.1em
  \fontdimen7\font=0.1em
  \hyphenchar\font=`\-
}

\renewcommand{\texttt}[1]{%
  \begingroup
  \ttfamily
  \begingroup\lccode`~=`/\lowercase{\endgroup\def~}{/\discretionary{}{}{}}%
  \begingroup\lccode`~=`[\lowercase{\endgroup\def~}{[\discretionary{}{}{}}%
  \begingroup\lccode`~=`.\lowercase{\endgroup\def~}{.\discretionary{}{}{}}%
  \catcode`/=\active\catcode`[=\active\catcode`.=\active
  \justify\scantokens{#1\noexpand}%
  \endgroup
}

\usepackage{fancybox,framed}

\begin{document}

\begin{frontmatter}
\title{Continuous Experimentation and the Cyber-Physical Systems challenge: An overview of the literature and the industrial perspective.}

\author[CTH]{Federico Giaimo\corref{mycorrespondingauthor}}
\cortext[mycorrespondingauthor]{Corresponding author}
\ead{giaimo@chalmers.se}
\author[CTH]{Hugo Andrade}
\ead{sica@chalmers.se}
\author[GU]{Christian Berger}
\ead{christian.berger@gu.se}
\address[CTH]{Chalmers University of Technology}
\address[GU]{University of Gothenburg}

\begin{abstract}
\textbf{Context:} New software development patterns are emerging aiming at accelerating the process of delivering value. One is Continuous Experimentation, which allows to systematically deploy and run instrumented software variants during development phase in order to collect data from the field of application. While currently this practice is used on a daily basis on web-based systems, technical difficulties challenge its adoption in fields where computational resources are constrained, e.g., cyber-physical systems and the automotive industry. 

\textbf{Objective:}
This paper aims at providing an overview of the engagement on the Continuous Experimentation practice in the context of cyber-physical systems.

\textbf{Method:} A systematic literature review has been conducted to investigate the link between the practice and the field of application. Additionally, an industrial multiple case study is reported. 

\textbf{Results:} The study presents the current state-of-the-art regarding Continuous Experimentation in the field of cyber-physical systems. The current perspective of Continuous Experimentation in industry is also reported. 

\textbf{Conclusions:} The field has not reached maturity yet. More conceptual analyses are found than solution proposals and the state-of-practice is yet to be achieved. However it is expected that in time an increasing number of solutions will be proposed and validated. 

\end{abstract}

\begin{keyword}
\CE \sep Cyber-physical Systems \sep Software Engineering
\end{keyword}

\end{frontmatter}

\section{Introduction}

Technology progresses at an ever-increasing pace: new ideas, new techniques, and new products are constantly being developed, threatening the industrial players with slower work methodologies. 
Product owners are thus forced to deliver value as quickly as possible in order to keep their edge. 
The software industry is a prime example of this trend, especially in some of its sub-fields, such as web-based software systems. 

Responding to this need for fast-paced value-centred software evolution, a number of practices have emerged with the goal of accelerating the processes around the development and deployment phases of software products' life cycle. 
Among them are some increasingly known and adopted Extreme Programming's Continuous Processes: Continuous Integration and Continuous Delivery/Deployment, which respectively advocate the integration of new code from developers' working copies into the main code tree often, ideally as soon as possible; 
and delivery or deployment of code to the products and systems as soon as it is integrated, where the difference between delivery or deployment consists in the presence or not of an automated deployment process. 
On top of these processes can sometimes sit an additional one, developed and adopted mainly in the context of web-based software-intensive systems, called \CE. 
It promises to introduce a real-world data feedback stream that can guide the development and evolution of existing and new features. 

\subsection{Background} 
\CE\ is a practice that is based on the idea of multiple A/B testing and relies on the fast release channels offered by Continuous Deployment. 
It results in having in a system or product the possibility to always run one or more different instrumented versions of the software in order to evaluate their performances, with the long-term goal of improving the system software via a series of incremental improvements validated from the field of use. 
More in detail, \CE\ differs from A/B testing since it allows to run A, B, and possibly more versions of the software on the same platform, while it executes its normal tasks. 
A more detailed description of this practice can be found in~\cref{sec:ce}.

Cyber-physical systems are \textit{integrations of computation and physical processes}~\citep{Lee08}, which means that these systems are immersed in the physical world and interact with it as the origin and/or result of their computation. 
This definition is quite broad and includes low-power and low-capabilities devices that are an important focus in some research and industrial areas, e.g.,~the Internet-of-Things. 
However, due to the computational and connectivity needs of a practice like \CE, the \cps\ that are referred to in the context of this work are those systems that are built with or that could accommodate adequate processing power and at least occasional connectivity capabilities.

Vehicles, which nowadays can contain more than a hundred \cps~\citep{URL_FUSE_Hiller}, could be considered as a sort of ``systems of \cps'' capable of fulfilling the aforementioned needs.
Additionally, many automotive companies are joining the trend of adding and improving their software capabilities to provide as much automation as possible to their customers. 
This means that they have the capability and the interest in exploring possible practices that can help a desirable evolution of their software functionality, for their customers. 
For these reasons, while the general interest is to enable \CE\ in \cps, the focus of this paper will be on the automotive systems.
This choice does not intend to exclude all other possible fields or systems, but before \CE\ could be applied in many of the current \cps\ sub-fields, there are still several technological challenges yet to overcome compared to the ones that the automotive systems face at the present development stage. 

\subsection{Motivation and Research Goal} \label{subsec:researchgoal}
While the use of \CE\ is a reality on web-based software-inten\-sive systems or smartphone apps, this is still far from true in the field of \cps. 

\begin{itemize}[label=$Research~Goal:$, leftmargin=*, labelindent=0em]
    \item \RGoal. 
\end{itemize}

The Research Goal was divided in the two following research questions and two different research methods were applied to answer them:

\begin{enumerate}[label=$RQ{\arabic*}:$, leftmargin=*, labelindent=0em]
    \item 
    \RQone
    \item 
    \RQtwo 
\end{enumerate}

To achieve the Research Goal and answer RQ1, a systematic literature review has been conducted to shed light on the link between the research and this field of application. 
The included primary studies are listed in \cref{tab:LRResults} and summarized in \cref{sec:resultsRQ1}. 
To answer RQ2, feedback from industrial practitioners was collected in two case studies conducted in two automotive companies. 
The results are described in \cref{sec:resultsRQ2}. 

\subsection{Contributions}
This article claims the following contributions: 
\begin{enumerate}[label=$C{\arabic*}:$, leftmargin=*, labelindent=0em]
    \item it summarizes the state-of-the-art of the research on \CE\ applied to the field of \cps; 
    \item it identifies the main challenges posed by \CE\ for automotive practitioners; and
    \item it identifies the main opportunities posed by \CE\ for automotive practitioners. 
\end{enumerate}
\subsection{Scope}
The scope of this work is the bond between the \CE\ practice and the \cps\ field, as opposed to studying the \CE\ practice in any possible field of adoption. 
This applies for both the research questions, but even more specifically for RQ2, where the scope is further focused on the automotive field. 
This choice is reflected by the keywords chosen in the literature analysis, where articles were included if they would express the link between these topics. 

\subsection{Structure of the document}
\cref{sec:ce} presents in details the concept of the \CE\ practice; 
\cref{sec:rm} describes the research strategy adopted in this study; 
\cref{sec:relwork} lists and summarizes related works; 
\cref{sec:res} reports the results of this work; 
\cref{sec:disc} discusses the results and their possible implications; finally, 
\cref{sec:confw} concludes this article, and describes possible directions for future efforts. 

\section{\CE}\label{sec:ce}
\begin{figure}[ht] 
  \begin{center}
  \includegraphics[width=.9\linewidth,trim=0cm 0cm 0cm 0cm]{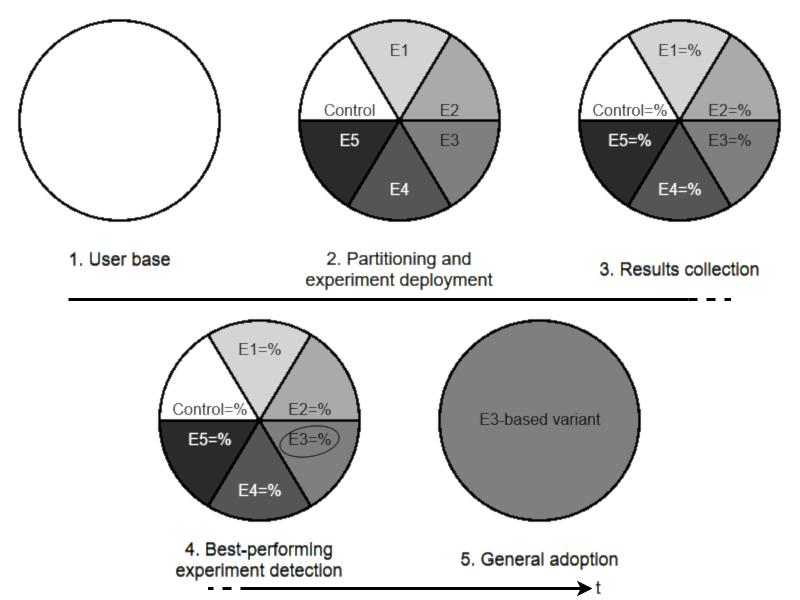}
        \caption{Timeline of the phases of the \CE\ process}
        \label{fig:CE}
  \end{center}
\end{figure}

Building upon the aforementioned Continuous methodologies, \CE\ is one Continuous practice that has recently gained momentum both in academia and among industrial practitioners in the field of web-based software-intensive systems.
The goal of \CE\ is to enable the product owner to steer the development of new functionality by measuring their impact in terms of real-world data with respect to one or more chosen metrics. 
This is achieved by deploying instrumented variants of the ``official'' software, the \textit{experiments}, through a process inspired by scientific experimentation that on the organizational side involves several figures and is composed by the following steps~\citep{FGMM17}:
\begin{enumerate}[label=Step~{\arabic*}:, leftmargin=*, labelindent=0em]
    \item One of the assumptions comprising the development plan for a product is chosen to be tested by the product owner;
    \item the data scientist receives the assumption and draws an experimentation plan comprising the details of the experiment to be run, the type of data that is expected and the analysis that will be performed on them.  
    In this step, a role knowledgeable about the system may be involved, complementing the data scientist's plan with their expertise on the system's capabilities;
    \item the developer receives the experimentation plan and implements it, while the release responsible roles deploy the experiment-primed software to the systems.
\end{enumerate}
From a more technical point of view, instead, the \CE\ process can be divided into the following phases, as shown in~\cref{fig:CE}: 
\begin{enumerate}[label=Phase~{\arabic*}:, leftmargin=*, labelindent=0em]
    \item the user (or system) base is defined, i.e.,~the set of users or deployed systems available for experimentation purposes; 
    \item the user base is divided in a number of significant partitions depending on the goal of the experiment, e.g.,~geographic localization, time of the day, etc. 
    To each of these partitions, except for a ``control partition'', an instrumented experiment is deployed. 
    Each experiment is a different variant of the software with a new or different functionality to be tested;
    \item the results from the experiments are collected and relayed back to the product owner and data scientist;
    \item the collected data is analyzed, possibly using statistical methods to remove noise and ignore human bias, and finally the best-performing experiment is identified;
    \item according to a fitting set of goal- and experiment-dependent metrics, the experiment that performed best is chosen for global adoption across the user (or system) base. 
\end{enumerate}

\section{Related Work}\label{sec:relwork}
Research on Continuous Experimentation is growing in time, as an increasing number of universities and companies acknowledge and study its potential. 
Some of these studies are relevant and related to the goal of the present work and their respective differences with this study will be outlined. 

\cite{FGMM17} defined their ``RIGHT'' model for \CE, an organizational model defining the tasks and artifacts that the different roles involved in the planning and implementation of a software product should manage in order to enable a smooth experimentation process. 
Their work however does not focus on the specific issues that \cps\ face, e.g.,~the resource constraints that may challenge the planned experiments or the impact that the presence of hardware components may have on the release of experiments. 

\cite{RR18} run a literature review to investigate what companies and what experiments are mostly performed in \CE. 
They mention attempting a pilot study in 2016, which did not find enough publications on the topic; independently from them, we also attempted a pilot study in that year, finding not enough published works as well.
Their findings draw a picture in which mainly big companies perform the most experiments, which are more often aimed at visual changes than algorithmic changes, the latter case being performed only with A/B experiments.
They also investigate which \CE\ research sub-topics are explored in literature, finding that experimentation infrastructure, challenges and statistical methods are the three most common ones. 
They mention but not focus primarily on the connection between \CE\ and \cps.

\cite{AF18} also run a literature review aimed at assessing the state of research on \CE\ and its main topics, contributors, and research types. 
They draw a picture of how \CE\ is spreading as a research subject to multiple venues and academic parties and similarly to \cite{RR18} finds a high presence of studies on statistical methods, infrastructure, and organizational topics applied to \CE. 
As well as the previous publication, they mention but do not focus on the connection between experimentation and \cps.

\cite{MBO18} run a literature review to identify challenges to the \CE\ process in \cps\ that were the object of a case study where they tried to identify possible solutions. 
While their work considers \CE\ and \cps, in their literature review the search query is generally on \CE\ and thus does not express the strong link with embedded systems that we are trying to highlight in the present work. 
 
\section{Research Method}\label{sec:rm}
\begin{figure}[ht]
  \begin{center}
  \includegraphics[width=.9\linewidth,trim=0cm 0cm 0cm 0cm]{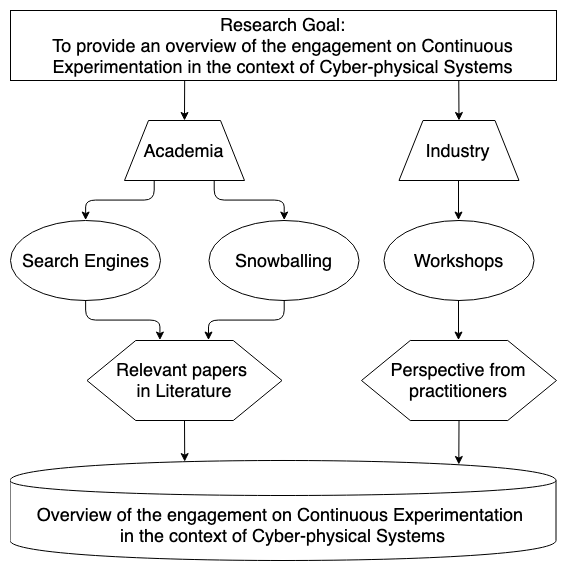}
  \caption{Research strategy highlighting the methodologies employed in this study}
  \label{fig:meth}
  \end{center}
\end{figure}

To assess the Research Goal and its research questions, a multi-method approach was devised in order to engage with different strategies for the research questions and gain a wider perspective on the topic. 
To answer RQ1 a systematic literature review was conducted, comprising both a query search and a snowballing phase~\citep{KBB15}. 
For RQ2 a multiple case study was performed in order to collect feedback from industrial practitioners~\citep{RH09}. 
An overview of the research strategy is shown in~\cref{fig:meth}.

\subsection{Literature Review (RQ 1)}
The first goal of this work is to present the state-of-the-art for the research on \CE\ in the field of \cps. 
To do so, a literature review was performed following the guidelines expressed by \cite{KBB15}. 
\subsubsection{Search strategy}
The search string was initially based on relevant related works that explored the literature with the aim of covering what progress has been made about the general study and adoption of \CE~\citep{RR18,AF18}. 
As our goal was to focus on the adoption of \CE\ in \cps, in the example of the automotive industry, we added to the search string relevant terms that would steer the scope of the search towards these specific sub-fields. 
Due to the novelty of the \CE\ practice and the lack of a globally accepted name in all the sub-disciplines that adopt this practice or variations thereof, many synonyms were added to the search string in order to obtain accurate results.
The majority of these search terms were used also by those related works that run comprehensive literature explorations. 
The problem posed by the presence of many synonyms in use for a certain practice or field does not appear for \cps, which is a more established research context with a widely accepted terminology. 
The final string is thus:

\begin{tabular}[t]{p{2cm} >{\ttfamily}p{6cm} p{2cm}}
(& ``continuous experimentation'' & OR \\
 & ``experiment systems'' & OR \\
 & ``controlled experiments'' & OR \\
 & ``controlled experimentation'' & OR \\
 & ``a/b testing'' & OR \\
 & ``a/b tests'' & OR \\
 & ``split testing'' & OR \\
 & ``split tests'' & OR \\
 & ``bucket testing'' & OR \\
 & ``bucket tests'' & OR \\
 & ``automated experiments'' & OR \\
 & ``automated experimentation'' & OR \\
 & ``live experiments'' & OR \\
 & ``live experimentation'' &) \\
 AND  \\(& ``cyber-physical'' & OR \\
 & ``embedded systems'' & OR \\
 & ``automotive'' &) \\
\end{tabular}

The search string was queried on the following databases: 
\textit{ACM Digital Library}, \textit{IEEE Xplore}, \textit{Scopus}, and \textit{Web of Science}, returning a total of 192 publications (results up do date as of October 2019). 
To improve the completeness of the search results, as suggested by~\cite{KBB15}, a set of 12 papers were used as the basis for a manual backwards snowballing phase, which added 211 publications.
These papers were chosen among the works included in past literature explorations~\citep{RR18,AF18,MBO18} due to their relevance in the field and to the scope and focus of this work. 

Successively the results were checked for duplicates. 
All results from the database and snowball search were collected in CSV format and a script comparing entries by publication title removed works which appeared more than once.

\subsubsection{Selection criteria}
The selection phase is performed after duplicates are removed, and is based on a set of selection criteria.
The selection criteria determine whether or not the retrieved studies are within the scope of this work. 
For this reason, the selection criteria are a fundamental building block of the study and require to be carefully defined in order to include all and only those publications which are relevant to the topic.
Two inclusion criteria were adopted and both had to be fulfilled by each study in order to be included. 
To judge whether the criteria were fulfilled, each study was read in its entirety. 
The criteria were: 
\begin{itemize}
    \item The study has a focus on \CE\ or A/B testing as a process, as opposed to a single test or experiment
    \item The study has a focus on the \CE\ process in the field of \cps, i.e., considering the resource limitations that ensue as opposed to \CE\ performed on web systems 
\end{itemize} 
A publication was instead excluded when any of the following exclusion criteria were met:
\begin{itemize}
    \item The publication is not in English 
    \item The publication is not peer-reviewed
    \item The publication is not a full paper (as opposed to a position paper, for example)
    \item The study is not a primary study 
\end{itemize} 

A summary of the results from the database search, backwards snowballing, duplicate removal and selection phases can be found in~\cref{tab:sms_results}. 

\begin{table}[H]
   \centering
   \caption{Search and snowballing results}
   \label{tab:sms_results}
   \begin{tabular}{|c|c|}
     \hline
     Database Name & Number of hits \\
     \hline
     ACM Digital Library & 39 \\
     IEEE Xplore & 13 \\
     Scopus & 125 \\
     Web of Science & 15 \\
     \hline
     Total database hits & 192 \\
     \hline
     \hline
     After snowballing & 403 \\ 
     \hline
     \hline
     Included publications & 8 \\
     \hline
\end{tabular}
\end{table}

To strengthen the confidence in the resulting included publications, a test-retest approach~\citep{KBB15} was employed, which means ``repeating (after a suitable time delay) some or all of the study selection actions'' in order to compare the outcomes. 
This was performed re-analysing the results obtained after the duplicates removal step in order to re-evaluate the selection criteria for each of the publications.

\subsection{Multiple case Study (RQ 2)}
In order to complement the systematic literature review and to additionally broaden the scope of the results, a multiple case study was devised to obtain empirical data from automotive industry representatives. 
This multiple case study extends the work reported by the authors in a previous article, where another multiple case study was performed adopting the same methodology, with the aim to extend, complement, and further validate the combined results~\citep{GAB19}. 
In this article the novel multiple case study is referred to as ``current multiple case study'' while the previously reported one as ``previous multiple case study''. 
The goal of the multiple case study was to ask the representatives the following working questions: 
\begin{enumerate}[label=${WQ\arabic *}:$,leftmargin=*,labelindent=0em]
\item What are the advantages that the \CE\ practice would bring in the context of autonomous driving with respect to their professional role in industry?
\item What are the challenges that the \CE\ practice would face in the context of autonomous driving with respect to their professional role in industry?
\end{enumerate}

\subsubsection{Format of the case study}
The case studies were conducted in a workshop format, each of them lasting between 1.5 and 2 hours, depending on the number of participants. During the workshops, one of the authors would lead it through its different phases, while the other authors would assist and take notes. 
The format was structured in four phases as follows:

\begin{enumerate}[label=Phase~{\Roman*}:~, leftmargin=\parindent,align=left,labelwidth=\parindent,labelsep=0pt] 
\item The workshop would begin with a presentation having the goal of establishing a common understanding and vocabulary of the Continuous practices, i.e., Continuous Integration, Continuous Delivery/Deployment, and \CE. 
This phase would last around 20 minutes; 
\item After the initial presentation, the participants were asked the two working questions about \CE, i.e.~WQ1 and WQ2. 
This phase would last around 30 minutes, during which the participants would individually write their answers, each different idea on a different note; 
\item The participants were asked to go through their notes to explain and clarify the meaning and reasoning behind each of them. 
Each note would then be placed next to others expressing similar ideas on a whiteboard, thus creating clusters around common ideas. 
This phase would last around 40 minutes; 
\item An infrastructure model for \CE\ devised for companies with web-based products~\citep{FGMM17} was introduced to the participants.
The aim was to start a discussion about the model and its criticalities if it had to be applied to the automotive industry. 
This phase would last around 15 minutes. 
\end{enumerate}

The format of these case studies was based on open questions focused on a structured topic, categorizing them as a series of semi-structured case studies~\citep{RH09}. 
This approach was chosen since it fits the exploratory and explanatory goal of the case study by promoting the participants to provide original feedback. 

Two automotive companies were chosen to run the described case study. 
Company A manufactures heavy-duty commercial vehicles. From this company 3 representatives joined the case study, 1 manager, 1 team leader and 1 engineer. Company B is an innovation center aimed at developing consumer vehicles capable of advanced capabilities. 
From this company 15 employees took part in our study, where 1 of them was a manager, 7 team leaders and 7 engineers.
To recruit participants, the authors reached out to their industrial network to sample key people who could be interested in the case study topic and/or could have contacts with other potentially interested parties. 
The overall variety of roles is considered a strengthening factor due to the increased diversity in points of view and resulting perspectives and discussions. 

Due to the strong connection between the present multiple case study and the previously reported one~\citep{GAB19}, some details about the composition of the latter will follow. 
The previous series of case studies involved four companies, adding to the two aforementioned novel cases. 
They comprised two automotive OEMs (Original Equipment Manufacturers) in this article named Companies C and D, a Tier-1 supplier named Company E, and an autonomous driving electric vehicle start-up company named Company F. 
The participants’ roles were: from Company C, 3 engineers, 1 team leader and 1 manager; from Company D, 2 engineers and 2 team leaders; from Company E, 1 engineer and 2 team leaders; lastly, from Company F, 1 engineer and 1 team leader.
To avoid biasing the participants of each case study, the themes and discussions resulting from any other case study were not disclosed.

\section{Results}\label{sec:res}

\subsection{Literature Review}\label{sec:resultsRQ1}

The analysis of the literature concerning \CE\ and \cps\ returned a research landscape that has not reached maturity and seems to be still searching for a definite direction forward. 
In fact, the selected articles, listed in \cref{tab:LRResults}, focused mostly on the depiction of the desire or needs of the industry~\citep{MBO18,OB14,OB13,EB12,BE12,Bosch12}, or on the identification of new methods and techniques~\citep{GB17,GBK17,EB12}. 
While the articles suggest new approaches and techniques, validation steps are rarely taken to verify whether the proposed approaches would yield the expected results in practice, which is interpreted by the authors as another byproduct of the novelty of the field.

\afterpage{
\thispagestyle{empty}
\begin{landscape}
    \centering
    \begin{table}  \vspace*{-3cm}\hspace*{-3.4cm} 
    \scriptsize
        \begin{tabularx}{25cm}{|p{3cm}|p{10cm}|p{5cm}|p{2cm}|p{2.81cm}|} \hline
            Authors & Title & Book & Year & Publisher \\ \hline
            Mattos D.I., Bosch J., Olsson H. H. & \href{https://doi.org/10.1007/978-3-319-91602-6_20}{Challenges and strategies for undertaking continuous experimentation to embedded systems: Industry and research perspectives} 
            & Lecture Notes in Business Information Processing & 2018 & Springer Verlag \\ \hline
            Giaimo F., Berger C., Kirchner C. & \href{https://doi.org/10.1007/978-3-319-65831-5_6}{Considerations about continuous experimentation for resource-constrained platforms in self-driving vehicles} 
            & Lecture Notes in Computer Science & 2017 & Springer Verlag \\ \hline
            F. Giaimo; C. Berger & \href{https://doi.org/10.1109/ICSA.2017.36}{Design Criteria to Architect Continuous Experimentation for Self-Driving Vehicles} 
            & Proceedings of the 2017 IEEE International Conference on Software Architecture (ICSA) & 2017 & IEEE \\ \hline
            Olsson, H. H., Bosch, J. & \href{https://doi.org/10.1109/SEAA.2014.75}{From Opinions to Data-Driven Software R\&D: A Multi-case Study on How to Close the 'Open Loop' Problem} 
            & Proceedings of the 40th EUROMICRO Conference on Software Engineering and Advanced Applications & 2014 & IEEE Press, New York \\ \hline
            Olsson, H. H., Bosch, J. & \href{https://doi.org/10.1007/978-3-642-39336-5_9}{Post-deployment Data Collection in Software-Intensive Embedded Products} 
            & Lecture Notes in Business Information Processing. LNBIP & 2013 & Springer, Heidelberg \\ \hline
            U. Eklund; J. Bosch & \href{https://doi.org/10.1109/WICSA-ECSA.212.38}{Architecture for Large-Scale Innovation Experiment Systems} 
            & Proceedings of the 2012 Joint Working IEEE/IFIP Conference on Software Architecture and European Conference on Software Architecture & 2012 & IEEE \\ \hline
            Bosch J., Eklund U. & \href{https://doi.org/10.1007/978-3-642-34026-0_3}{Eternal embedded software: Towards innovation experiment systems} 
            & Lecture Notes in Computer Science & 2012 & \\ \hline
            Bosch, J. & \href{https://doi.org/10.1007/978-3-642-30746-1_3}{Building Products as Innovation Experiment Systems} 
            & Lecture Notes in Business Information Processing. LNBIP & 2012 & Springer, Heidelberg \\ \hline
        \end{tabularx} 
    \caption{Results from the literature investigation}
    \label{tab:LRResults}
    \end{table}
\end{landscape}
}

The main findings of the selected articles can be summarized as follows. 
\cite{MBO18} identify a number of challenges in adopting \CE\ on \cps\ from both the academic and industrial contexts, they also provide a set of possible strategies to overcome these challenges as suggested by industrial representatives in their case studies. 
\cite{GBK17} focus on the issue of the scarcity of the necessary computational resources in \cps\ that would benefit from the implementation of \CE, suggesting three possible execution strategies to overcome it.
A number of necessary design criteria for \cps\ that are expected to run \CE\ techniques are identified by \cite{GB17}; among the criteria they propose characteristics that the software development process should have in order to facilitate the adoption of the practice. 
The work by \cite{OB14} and \cite{OB13} provides process models and techniques that focus on the collection of feedback data from the products and customers in the post-deployment phase of the software development of the product. 
An architecture for experiments called ``innovation experiment systems'' is proposed by \cite{EB12}, \cite{BE12}, and \cite{Bosch12}; additionally they run case studies involving the proposed architecture performing A/B tests on an automotive infotainment system and in the context of a company providing software-as-a-service in the context of connected embedded systems. 

The studies resulting from the literature review are summarized singularly in the following tabs.

~

\noindent\framebox{\parbox{\dimexpr\linewidth-2\fboxsep-2\fboxrule}{
\textbf{Title:} 
Challenges and Strategies for Undertaking Continuous Experimentation to Embedded Systems: Industry and Research Perspectives~\citep{MBO18}

\textit{Scope:} 
Continuous experimentation and the challenges and requirements that embedded systems companies have to run experiments in their systems.

\textit{Research Goal:} 
Exploring the challenges posed by the adoption of continuous experimentation in embedded systems.

\textit{Methodology:} 
Literature review and multiple case study based on interviews and workshop sessions.

\textit{Contributions:} 
Challenges from both literature and industrial experience, and possible strategies to overcome them.

\textit{Conclusions:} 
The set of identified challenges are presented with a set of strategies and solutions to overcome them; additionally, open challenges and the need for new tools are pointed out. 

\textit{Threats to Validity:} 
Scope of the literature review, generalization of the collected challenges. 

}}

~

~

\noindent\framebox{\parbox{\dimexpr\linewidth-2\fboxsep-2\fboxrule}{
\textbf{Title:} 
Considerations About Continuous
Experimentation for Resource-Constrained
Platforms in Self-driving Vehicles~\citep{GBK17}

\textit{Scope:} 
\CE\ and its technical challenges on \cps\ on the example of automotive systems.

\textit{Research Goal:} 
To assess the scarcity of resources that could disrupt or prevent the adoption of \CE\ on \cps.

\textit{Methodology:} 
Exploratory study, design science.

\textit{Contributions:} 
Three technical strategies to circumvent the physical limitations of \cps\ with the aim of enabling \CE; description of software architecture capabilities that would enable them.

\textit{Conclusions:} 
The execution strategies are presented together with their prerequisites in the software infrastructure. 

\textit{Threats to Validity:} 
Validation underway but not reported.

}}

~

~

\noindent\framebox{\parbox{\dimexpr\linewidth-2\fboxsep-2\fboxrule}{
\textbf{Title:} 
Design Criteria to Architect Continuous
Experimentation for Self-Driving Vehicles~\citep{GB17}

\textit{Scope:} 
Architectural needs for \CE\ on complex \cps\ such as self-driving vehicles. 

\textit{Research Goal:} 
The goal of the paper is to find properties of the software architecture and process required to enable \CE\ for a complex cyber-physical system.

\textit{Methodology:} 
Literature analysis and design science.

\textit{Contributions:} 
List of properties or features that a software architecture should provide in order to enable \CE\ on \cps. 

\textit{Conclusions:} 
The study concludes underlining that \cps\ can benefit from \CE, although technical challenges still exist that impede a widespread adoption. 

\textit{Threats to Validity:} 
The scope of literature exploration, focus not on safety considerations
}}

~

\noindent\framebox{\parbox{\dimexpr\linewidth-2\fboxsep-2\fboxrule}{
\textbf{Title:} 
From Opinions to Data-Driven Software R\&D~\citep{OB14}

\textit{Scope:} 
Embedded software companies.

\textit{Research Goal:} 
The goal of this paper is to find mechanisms that help companies confirm that the product features they prioritize are of value for customers. 

\textit{Methodology:} 
Multiple case study.

\textit{Contributions:} 
A process model to guide the companies to adopt practices that return a feedback from their customers. 

\textit{Conclusions:} 
The model enhances productivity due to its focus on customer validation of the companies' efforts. 

\textit{Threats to Validity:} 
Construct validity for the topics in the case studies, generalization of the findings.

}}

~

~

\noindent\framebox{\parbox{\dimexpr\linewidth-2\fboxsep-2\fboxrule}{
\textbf{Title:} 
Post-deployment Data Collection in Software-Intensive Embedded Products~\citep{OB13}

\textit{Scope:} 
Companies involved in large-scale development of embedded 
products. 

\textit{Research Goal:} 
To provide an overview of post-deployment data usage in the embedded products' industry.

\textit{Methodology:} 
Multiple case study.

\textit{Contributions:} 
An inventory of techniques used for customer involvement and customer feedback collection before, during and after product development. 
It also presents opportunities for more effective product development and evolution in the post-deployment phase of software development. 

\textit{Conclusions:} 
The authors highlight limitations in the research and practice of post-deployment data collection aimed at the improvement and innovation of the existing deployed systems, as opposed to troubleshooting.

\textit{Threats to Validity:} 
Construct validity for the topics in the case studies.
}}

~

~

\noindent\framebox{\parbox{\dimexpr\linewidth-2\fboxsep-2\fboxrule}{
\textbf{Title:} Architecture for Large-Scale Innovation Experiment Systems~\citep{EB12}

\textit{Scope:} 
Embedded systems domain. 

\textit{Research Goal:} 
The goal of the paper is to define principles for the architecture of large-scale experiments. 

\textit{Methodology:} 
Design science, case study.

\textit{Contributions:} 
Theoretic infrastructure for experiments on embedded systems.

\textit{Conclusions:} 
The authors proposed an architecture for experiments called ``innovation experiment system'' and studied an industrial case adopting the architecture in an A/B test.

\textit{Threats to Validity:} 
Proposed architecture may not be complete, validation on only one case study presented.
}}

~

~

\noindent\framebox{\parbox{\dimexpr\linewidth-2\fboxsep-2\fboxrule}{
\textbf{Title:} 
Eternal Embedded Software:
Towards Innovation Experiment Systems~\citep{BE12}

\textit{Scope:} 
Long-lived embedded systems.

\textit{Research Goal:} 
To introduce the notion of ``innovation experiment system'' and to apply it to the context of long-lived embedded systems. 

\textit{Methodology:} 
Exploratory study, case study.

\textit{Contributions:} 
The contribution of the paper is a discussion of the concept of innovation experiment systems, exploring the architectural implications of such systems, and it illustrates a case study concerning an infotainment system in the automotive industry.

\textit{Conclusions:} 
The proposed architecture for experimentation can help embedded systems to evolve and respond to changing context and requirements. 

\textit{Threats to Validity:} 
Validation on only one case study is presented. 

}}

~

~

\noindent\framebox{\parbox{\dimexpr\linewidth-2\fboxsep-2\fboxrule}{
\textbf{Title:} Building Products as Innovation Experiment Systems~\citep{Bosch12}

\textit{Scope:}
This paper looks at the evolution of the development process of Software-as-a-Service (SaaS) solutions and software-intensive embedded systems.

\textit{Research Goal:}
To address the application of experimentation, ranging from optimization of existing features to the development of new features and products.

\textit{Methodology:}
Case study.

\textit{Contributions:} 
A systematization of the proposed ``innovation experiment system'' approach to software development for connected systems, and the illustration of the model using an industrial case study.

\textit{Conclusions:}
The authors note that the traditional development approaches are being replaced by new ones, focusing on factors like continuous evolution and utilization of user data. 

\textit{Threats to Validity:} 
Proposed systematization may not be complete, validation on only one case study presented. 
}}

~

To further summarize the answer to RQ1: 

\noindent\doublebox{\parbox{\dimexpr\linewidth-2\fboxsep-2\fboxrule}{
The majority of studies have a high-level approach to the topic, mostly describing what challenges \CE\ faces if applied on \cps; many of these are empirical studies, aimed at gathering data from practitioners; only a minority of articles are design studies proposing solutions to the challenges that \CE\ faces in the field of \cps. }}

\subsection{Multiple case study}
\label{sec:resultsRQ2}

In this section the resulting data from the multiple case studies are collected. 
The notes written by the participants were analysed and grouped in semantic clusters, resulting in the two-level lists that follow, one for the reported Advantages and one for the Challenges. 
In both description lists, each high-level theme (in boldface characters) represents a cluster, which contains one or more detailed items (in italic characters), representing the single ideas put forward by the participants. 
Due to the complex nature of the problem, some items may be related to each other due to fundamental topics and issues that span and affect multiple thematic aspects. 
The connection between which item was mentioned in which companies, including the data from both the current and previous multiple case study, is shown in~\cref{tab:AV,tab:Ch}. 

\subsubsection{Advantages description list}

\textbf{Safety:}
Software-enabled auxiliaries to basic functions like braking and steering could reduce the risk of dangerous situations occurring during the products operational life. 
With a constant loop of experimentation and updates, the robustness of the software in unforeseen or perilous events would increase over time and therefore improve the overall safety of the system.

\begin{itemize}
\item \textit{Monitoring.}
With the capability of communicating remotely with the products, it may be possible to find out product issues in a faster way. 
The monitoring could be employed not only for the software aspects but also for the mechanical integrity of vehicles, allowing product owners to be aware of and mitigate the impact of the wear and tear in their products. 

\item \textit{Reliability.} 
Constant monitoring could result in a better localization of errors and miscalculations, leading to more robust and reliable products overall.

\item \textit{Active/passive safety possibilities.}
Taking advantage of fast testing opportunities and time-to-market cycles, Continuous experimentation would allow new possibilities for active and passive safety functionality, i.e.~techniques to improve safety respectively before and during an accident. 
Novel possibilities and techniques can be experimented and improved based on the data collected from the field. 

\item \textit{Traffic prediction.}
With the constant transfer of sensor data to the headquarters, engineers can develop functionalities that are based on an always-improving representation of the world. 
Such amounts of data allow for better prediction of traffic behavior, which in turn improves road safety.

\end{itemize}

\begin{table}[H]
    \small
    \centering
    \begin{tabular}{>{\centering\arraybackslash}m{3cm}|>{\centering\arraybackslash}m{4cm}|>{\centering\arraybackslash}m{2cm}|>{\centering\arraybackslash}m{2cm}}
         Category & Advantage & Companies in current case study & Companies in previous case study \\\hline\hline
         \multirow{4}{*}{Safety} & Monitoring & B & D , E , F \\\cline{2-4}
         & Reliability & B \\\cline{2-4}
         & Active/passive safety opportunities & B \\\cline{2-4}
         & Traffic prediction & B \\\hline
         \multirow{3}{*}{Speed} & Faster data collection & A & C \\\cline{2-4}
         & Faster functionality feedback & B \\\cline{2-4}
         & Faster time-to-market & A , B & C , D , E , F \\\hline
         \multirow{3}{*}{Quality} & Customer satisfaction & B & D , E , F \\\cline{2-4}
         & Improved quality & B \\\cline{2-4}
         & Better world understanding & A \\\hline
         \multirow{4}{*}{Opportunities} & Reducing long-term costs &B \\\cline{2-4}
         & Monetization of data & B \\\cline{2-4}
         & Testing of 'bold' ideas & A \\\cline{2-4} 
         & Improving future solutions' design & A \\\hline
         - & Mechanical integrity & & E , F \\\hline
         - & Easier testing & & C , D , E \\\hline
         - & Energy efficiency & & F \\\hline
         - & Real-world data usage & & C , D , E \\\hline
         - & Incremental delivery & & E \\\hline
         - & Fleet view & & C \\\hline 
    \end{tabular}
    \caption{Perceived Advantages in the \CE\ practice and the companies raising each point. The first column contains the category of each Advantage, which is named in the second column, the third contains the companies that mentioned the item during the current multiple case study, and the fourth contains the companies that mentioned the item during the previous multiple case study, if any}
    \label{tab:AV}
\end{table}

\begin{table}[H]
    \small
    \centering
    \begin{tabular}{>{\centering\arraybackslash}m{3cm}|>{\centering\arraybackslash}m{4cm}|>{\centering\arraybackslash}m{2cm}|>{\centering\arraybackslash}m{2cm}}
         Category & Challenge & Companies in current case study & Companies in previous case study \\\hline\hline
         \multirow{2}{*}{Safety} & Impact measurements & B & D , E , F \\\cline{2-4}
         & Responsibility & B \\\hline
         \multirow{2}{*}{Security} & Data protection and privacy & A , B & C , D , E , F \\\cline{2-4}
         & Misuse of data & B \\\hline
         \multirow{3}{*}{Quality assurance} & Complexity of software and operations & B \\\cline{2-4}
         & Data quality & B \\\cline{2-4}
         & Validation and verification & A & C , D , E , F \\\hline
         \multirow{4}{*}{Costs} & Costs for experiment data management & A , B & C \\\cline{2-4}
         & Regulation changes & B \\\cline{2-4}
         & Costs of experiments & A \\\cline{2-4}
         & Tools to enable/support experimentation & A , B & C \\\hline
         \multirow{3}{*}{DevOps} & Data and configuration management & A , B & C , D \\\cline{2-4}
         & Software and hardware infrastructure & B \\\cline{2-4}
         & Global engineering & B \\\hline
         Hardware & Resource constraints & A & C , D , E \\\hline
         - & Fallback Plan & & F \\\hline
         - & Regulations & & C , D , E , F \\\hline
         - & Versioning & & C , E \\\hline
         - & Performance & & E , F \\\hline
         - & Remote execution & & E \\\hline
         - & Testing & & C \\\hline
         - & Heterogeneity & & D , E \\\hline 
    \end{tabular}
    \caption{Perceived Challenges in the \CE\ practice and the companies raising each point.}
    \label{tab:Ch}
\end{table}

\textbf{Speed:}
It has been reported that one crucial benefit in achieving \CE\ is the resulting increase in the speed of software development, testing, and release processes. 

\begin{itemize}
\item \textit{Faster data collection.}
With a constant connection between the headquarters and the vehicle, interesting data could be collected on demand, allowing for fast and ad hoc analysis of system behavior. 
Instead of collecting data from controlled tests on test tracks, the OEMs would benefit from the real-world system usage thanks to the Over-The-Air (OTA) connection. 

\item \textit{Faster functionality feedback.}
Faster data collection also allows for faster feedback from the users about the products' functionality. 
Preferences in terms of often-used or seldom-used functions can be detected and used to help the development process. 

\item \textit{Faster time-to-market.}
Updates would equally be fast-paced given that two-way OTA connectivity is established. 
Software could be updated regularly and without manual delivery of new versions. 
It could be faster to fix issues and improve the software establishing a more dynamic life-cycle. 
Instead of prototyping and running typical acceptance testing with a reduced number of users, the acceptance could be measured from real-world scenarios as fast as the data can be transmitted from the products back to the headquarters. 
Furthermore, simulations of the world can be enhanced thanks to the increasing amounts of data collected in the real world. 

\end{itemize}

\textbf{Quality:}
Quality has shown to be a concern of great importance in the adoption of \CE. 
The changes in the software process must not negatively affect the already conquered quality of the software and the customers' satisfaction. 

\begin{itemize}
\item \textit{Customer satisfaction.}  
The functionality of the software can be reassessed using statistics about the regular usage of the systems. 
The customers' preferences would be captured and implemented into the system through updates, improving customer satisfaction. 

\item \textit{Improved quality.}  
Acting on the constant feedback from the internal software performances and the interaction between customers and products, the overall quality of the products is expected to improve. 
Further, feedback on the performance of specific functions can be collected and assessed quickly. 

\item \textit{Better understanding of the world.}
Since experiments can be done at a larger scale than what is currently possible, the amounts of data would also increase. 
The systematic analysis of this large amount of data \textit{upstreaming} from the products would result in a better representation of the world to the benefit of simulations and future development efforts. 

\end{itemize}

\textbf{Opportunities:}
Some opportunities were pointed by the practitioners in the case of adoption of \CE.

\begin{itemize}
\item \textit{Reduced costs in the long run.}
Incremental and constant delivery of functionalities based on real-world scenarios and data may decrease the cost of development in the long run, or decrease the risk of deploying faulty software which is expensive to correct. 

\item \textit{Monetization of data.}
Data collected from the field could be monetized to third parties according to the owner company's business goals.

\item \textit{Possibility to test bold ideas.}
Companies would have the opportunity to test bold ideas in real-world usage scenarios, instead of simulations or test tracks. 
This can give more freedom to the developers and enable them to find novel and potentially better approaches in solving issues or improving functionality. 

\item \textit{Improving future solutions' design.}
Better design and development of new solutions in the future can be achieved thanks to better understanding of the real-world in combination with detailed understanding of how the products are actually used. 

\end{itemize}

\subsubsection{Challenges description list}

\textbf{Safety:}
Perhaps the biggest concern is how to ensure the safety of experimental versions of the system. 
Changes in the code base might negatively impact critical safety features. 
A robust strategy for obtaining a full understanding of such impacts is needed in order to deploy safe software to the vehicles. 
Safety requirements must also be guaranteed employing redundancy of critical hardware and software. 

\begin{itemize}
\item \textit{Impact measurements.}
Safety-critical applications strive for consistency and means to measure the impact of changes to the code base. 
Such measurements must occur before the deployment phase, which means that the real impact of changes would not be entirely under control. 
This scenario poses a challenge to testing, for instance, experiments that may affect the control of the vehicle. 

\item \textit{Responsibility.}
In case of accidents involving systems running experiments, the responsibility may be up for discussion. 
In addition to the governmental regulations, there might be margins for interpretation upon eventualities. 

\end{itemize}

\textbf{Security:}
Another major concern discussed in the workshops was the aspect of information security. 
Safely storing and transmitting user data or software requires the implementation of robust security mechanisms. 

\begin{itemize}
\item \textit{Data protection and privacy.}
Since both user information and experimental algorithms will move to and from the vehicle, one important concern would be the security of such communications. 
The integrity of the transmission must be preserved through security mechanisms that reduce the risk of interception, impersonation, or tampering by third-party entities. 
Furthermore, corporate secrecy might also play a role, since experiments will be embedded in products. 
Finally, companies may have to anonymize the data collected from the vehicles to comply to strong privacy laws such as the European General Data Protection Regulation (GDPR).

\item \textit{Misuse of data.}
Personal data belonging to customers could be misinterpreted or used for improper purposes by the companies themselves.
\end{itemize}

\textbf{Quality assurance:}
\CE is expected to bring an increase to software quality due to the inherent learning opportunities it offers to the developers. 
However, a number of topics were raised that could challenge the rise in quality, as follows.

\begin{itemize}
\item \textit{Complexity of software and operations.}
Running various instances of the software systems increase the complexity of the system. 
Multiple instances of the same software, including experimentation modules, also increase the complexity of the operations. 
Handling such increase in complexity poses an important challenge to \CE\ practitioners.

\item \textit{Data quality.}
When data arrives at the development site, collected from the field, there may be cases in which it is not clear how much it can actually be trusted as representative of reality. 
It could be the case that for determined purposes the data are not consistent enough to draw significant conclusions.

\item \textit{Validation and verification.}
Also connected with the measurement of impacts, companies implementing \CE\ must develop and assess robust procedures that allow for proper validation and verification of the software before it reaches the target systems.
\end{itemize}

\textbf{Costs:}
Industrial practitioners are concerned with the costs involved in implementing \CE. 
In particular, a novel hardware infrastructure would be necessary to accommodate software instances and transmit data to/from the target systems.

\begin{itemize}
\item \textit{Data management.}
Managing large amounts of data demands costs that must be accounted for when implementing \CE. 
For instance, the costs for storage, analysis, and transmission of the data collected by the systems in the fleet.

\item \textit{Regulation changes.}
Regulatory changes might be unforeseen and demand fundamental changes in the business model. 
The impact on research and development is typically high with respect to costs and the implementation of new processes.

\item \textit{Costs of experiments.}
There might be additional hidden costs tied to the design and implementation of experiments in the \CE\ fashion, which may not be foreseeable until the specific experiments are designed. 

\item \textit{Tools to enable/support experimentation.}
There would be inherent costs to implementing and/or buying hardware, software, and analytical tools to enable or support \CE\ in a large-scale organization, or to examine the results in a scientifically sound fashion. 
\end{itemize}

\textbf{DevOps:}
Practitioners mentioned challenges related to DevOps processes when possibly implementing \CE.

\begin{itemize}
\item \textit{Data and configuration management.}
Collecting, structuring, and analyzing data obtained from the field would become an integral part of the development process. 
The large amount of data collected could pose a managerial challenge in \CE. 
To reduce the load for the systems in the fleet, practitioners may need to decide what data would be relevant for collection and analysis and what instead could be discarded. 

\item \textit{Software and hardware infrastructure.}
In the context of experimental applications, the process would require both a software and hardware infrastructure to realize \CE. 
From the necessary software stack to run applications on the vehicle, to the required hardware for executing extra portions of code.

\item \textit{Global engineering.}
Several automotive projects contemplate global products, which adds an additional layer of complexity on the data collection. 
As an example, what could be a preference for a certain geographic market could be less desirable or impossible to achieve in another.
\end{itemize}

\textbf{Hardware:}
Additional hardware would most likely be needed to accommodate \CE\ in the existing systems. 
In some domains, such as the automotive field, adding weight and requiring extra space in the vehicle for additional equipment might be a crucial constraint. 
\begin{itemize}
\item \textit{Resource constraints.}
Highly resource-constrained computational units like those generally employed in the automotive field could potentially limit in a significant way the options for experimentation, making the addition of more advanced computational units necessary. 
\end{itemize}

\subsubsection{Complementing our previous study}
The reported results extend and complement aspects that emerged in a previous multiple case study performed by the authors~\citep{GAB19}. 
In that study the same categories (in boldface characters) emerged for both Advantages and Challenges, with the exception of the ``Sustainability'' item in the Advantages section. 
A number of additional subcategories (in italic characters), however, were not mentioned in the discussions during the latest case studies and, hence, did not appear in the above description lists. 
These items in the Advantages list were: 

\textbf{Safety}
\begin{itemize}
    \item \textit{Mechanical integrity}. 
    Constant monitoring results in a slower wear and tear of mechanical components by interpreting situational/behavioral states of the system. Once identified, wear-prone situations could be avoided.
    \item \textit{Easier testing}. 
    Field testing on the fly makes it easier to detect bugs, and with the constant feedback it would be easier to find relevant test cases for the system.
\end{itemize}

\textbf{Sustainability}
\begin{itemize}
    \item \textit{Energy efficiency}. 
    Unused functionalities can be disabled to reduce energy consumption. The data resulting from a constant monitoring of the hardware's energy consumption can also be used to improve energy efficiency. 
\end{itemize}

\textbf{Opportunities}
\begin{itemize}
    \item \textit{Real-world data usage}. 
    Learning from data enables research and improvements of both the process and the product. Further, the collected data can be analyzed and/or sold as services.
    \item \textit{Incremental delivery}. 
    Large and complex functions can be delivered step-by-step. Certain functions can be implemented in a bare-minimum fashion and updated and extended at a later time. 
    \item \textit{Fleet view}. 
    Companies may have the opportunity to obtain a comprehensive view of the behavior of their products based on the collected data from the fleet. 
\end{itemize}

Finally, the non-repeated items in the Challenges list were: 

\textbf{Safety}
\begin{itemize}
    \item \textit{Fallback plan}. 
     In case of failures, a fallback plan must always be ready. 
     With multiple versions of the software deployed, this solution demands a robust versioning system that allows safe rollback in case of emergencies.
     \item \textit{Regulations}. 
     Complying with strict governmental regulations (e.g.,~those in the automotive domain) can be a challenge in the case of experimental software. 
\end{itemize}

\textbf{DevOps}
\begin{itemize}
    \item \textit{Versioning}. 
Developers must acknowledge/monitor versions that are deployed. Different configurations of the same software may be deployed and running on different vehicles.
\end{itemize}

\textbf{Quality assurance}
\begin{itemize}
    \item \textit{Performance}. 
    Running various instances of the software can be very demanding to the automotive hardware, which is typically resource-constrained. 
    \item \textit{Remote execution}. 
    Data collection and important updates could be at risk of not occurring due to poor, faulty, or non-existing network connections. 
    \item \textit{Testing}. 
    Since most of the testing in the automotive industry is done manually, this stage currently involves very high costs. 
    It could be hard to test experimental software before it reaches the target systems.
\end{itemize}

\textbf{Hardware}
\begin{itemize}
    \item \textit{Heterogeneity}. 
    Systems with different hardware specifications pose a challenge in ensuring that new software versions are supported by the available hardware platforms with their different setups. 
\end{itemize}

To further summarize the answer to RQ2:

\noindent\doublebox{\parbox{\dimexpr\linewidth-2\fboxsep-2\fboxrule}{The main advantages deriving from the adoption of \CE\ on \cps\ emerged to be the reduction of the development time for new software, together with the possibility to better monitor the systems, and a possible increase of customer satisfaction; 
on the other hand, the main challenges are considered to be the privacy issues linked to the data resulting from experiments, the need to foresee the impact of the software changes that are pushed to vehicles, and the need to ensure validation and verification of the software that will run on vehicles, including the experimental software.}}

\section{Discussion}\label{sec:disc}

\subsection{State-of-the-art of research on \CE\ for cyber-phys\-i\-cal systems}
The \CE\ practice has been recently investigated in literature as noted also in \citep{AF18}, although in the context of \cps\ this has happened with a quite limited number of strategies and studies. 
As it emerged from the presented systematic literature review, the majority of studies have a high-level approach. 
This means that they try to tackle from a more conceptual point of view the difficulty of applying this practice to a new field which faces different challenges than the field from which \CE\ originates. 
Many of these studies are observational, in which a case study is run to gather feedback from practitioners or to analyze whether certain hypotheses are met in practice~\citep{MBO18,OB14,OB13,EB12,BE12,Bosch12}. 
A minority of articles are instead design studies trying to draft possible solutions to the technical hurdles opposing the adoption of \CE\ on \cps~\citep{GB17,GBK17,EB12}. 
This unbalance towards more theoretical studies is assumed by the authors to be a direct effect of the relative novelty of the practice in object in the field of \cps: 
in time it is foreseeable to see an increase in more technical studies facing and overcoming the challenges identified in this more investigative initial period. 

An interesting comparison can be drawn between these results and the ones reported in literature investigations performed in related studies.
Both \cite{RR18} and \cite{AF18} report that in \CE\ literature, i.e., \CE\ applied not only to the \cps\ field, the studies on solution proposal or validation studies are the minority, while the majority of studies performed experience reports and evaluation research. 
Topic-wise the most common focuses among experience reports and evaluation research appear to be the challenges that the adoption of \CE\ faces and the software infrastructure in place to enable it; in the case of solution proposals or validation studies, instead, the most common topics were statistical methods and the design of experiments. 
These results conform with the present systematic literature study, e.g., the majority of studies are evaluation research, which includes case studies, and only the minority are solution proposals studies; the topics are quite similar as well, with challenges to \CE\ and software infrastructure being central themes in many cases, as opposed to statistical analysis, which in this case was not mentioned in the selected articles. 

\subsection{Automotive practitioners' feedback on \CE\ in the \cps\ context}
Both companies in the current multiple case study highlighted that the most clear advantage of adopting \CE\ would be a reduction of the development time for new software. 
Many other desirable capabilities and effects were brought up but interestingly not by representatives in both companies. 
Some of the reported ideas had been stated also by other companies' participants in the previous multiple case study, e.g.,~the possibility to monitor the vehicle in terms of maintenance needs, the quicker data collection possibilities, and the quality feedback given to the software by the users. 
New items also emerged, such as the possibility to predict traffic patterns over time, the monetization of the collected data, or even the possibility to test bolder ideas than with the current tools and processes -- 
although the practicality of this last point is quite dependent on the context of the ideas themselves, since safety consideration must be taken into account before developing experiments. 

Drawing a comparison with the results of the previous multiple case study, there exist a relatively small overlap between the items in the \textit{Advantages} list collected in the current case studies and the ones collected during the previously reported case studies, meaning that the remaining items and considerations were not repeated. 
This could either hint at the broad spectrum of possible applications that the \CE\ practice could enable in this field, or at the uncertainty of the practitioners about what would actually be possible and what would not, or possibly a combination of these two elements. 
Considering the relative novelty of the practice in this context, however, a certain degree of spread in the collected ideas is not surprising. 

Moving the focus on to the \textit{Challenges} items, it is possible to observe that, similarly to what happened with the \textit{Advantages}, there are some items which were repeated and others that were unique for each single case study. 
Notably, the companies of the current case study agree that important challenges are, among the others, (i) ensuring customers' data protection and (ii) the management of the experimental data, together with (iii) the associated costs. 
Less unanimous but fruitful nonetheless were the discussions about interesting items such as the challenge of elaborating meaningful experiments, the problem of assigning responsibility in case of accidents, the trustworthiness of the collected data, or even the challenge of managing experiments running on systems distributed on a wide geographic scale, where cultural differences may have a bigger impact on the results than expected.  

Comparing the previous multiple case study with the current one, some items did not emerge in the latest cases, e.g.,~the presence of a fallback plan in case of failures during the experimentation process, or the risks associated with needing to exchange data with a product in an area where it cannot establish a successful connection, or the challenge to manage heterogeneous hardware configuration in different product families. 
The overlap between the \textit{Challenges} items in the two multiple case studies shows to be higher than what was seen in the \textit{Advantages} list, meaning that more agreement is found when discussing obstacles to the adoption of \CE\ in this field.

\subsection{Overview of \CE\ on \cps, with a focus on the automotive field} 

The aim of this work is to provide an overview of the engagement in \CE\ in the context of \cps, in the example of the automotive field. 

From the literature study it emerged that most articles either focus on the issue of enabling \CE\ on \cps\ and on preparing the software infrastructure from a conceptual standpoint, or report case studies where companies move initial steps towards the adoption of experimentation as a way to improve processes and products. 
Fewer studies try instead to propose solutions to specific technical issues. 
The predominance of studies on the challenges hint at a field which is still in its infancy, where important issues are still unsolved and hurdling prospective scholars and practitioners. 

Similar considerations can be drawn analyzing the findings of the conducted empirical studies. 
This different approach resulted in fact in a series of broad positive expectations and even broader issues that are currently preventing the adoption of experimentation in the industrial context, at least for what concerns automotive \cps. 
This means that a solid state-of-practice has not yet been established, as many interested parties are still working towards achieving a functional methodology to apply this practice. 

The accordance between the results from the literature and the empirical study highlights that a number of challenges still need to be solved or circumvented before \cps\ could reap in a systematic way the same benefits that the \CE\ practice has brought to the web-based software-intensive systems applications.
More specifically, the main issues that need to be solved appear to be firstly the ones connected to regulatory issues, both regarding what software can be run on vehicles and what can be done with respect to the privacy of the collected data; and secondly the processes around developing for and applying the practice, which encompasses the issues posed by the low computational resources available on vehicles, the tools needed to run \CE, and the way to organize the data and software configurations. 
To achieve this, it is desirable to see a future increment in design studies and solution evaluation studies that could devise and test architectures and technical solutions to bring forward the field. 

\subsection{Threats to Validity}
A first threat to the validity of this study is the possibility during the literature exploration to have not found all the articles that are relevant to our topic. 
To reduce this chance the investigation was conducted submitting the search query to multiple search engines and complementing the results with a snowballing phase. 

Moreover, threats to the validity of the literature exploration results may lie in the selection process. 
To increase the trustworthiness of the selection outcome, a test-retest approach was employed. 
This approach ``can be interpreted as being for the researcher to perform such tasks as \textit{selection} and \textit{data extraction} twice, with these being separated by a suitable time interval, and to check for consistency between the two sets of outcomes''~\citep{KBB15}.

A threat to the construct validity of the multiple case study results is the possibility that the first phase of the case studies, which included a presentation, had biased the participants' answers to the workshop questions. 
To limit the impact of this threat, the authors tried as much as possible to avoid content and examples that could influence in a certain direction the participants' thinking but to establish a common vocabulary for the workshop. 

A threat to the internal validity of the conclusions is the absence of data triangulation in the multiple case study, which involves running more than one time the same workshops in the same format to confirm the findings. 
The data triangulation was made impossible by the limited availability of the industrial representatives that joined the case studies. 

A threat to the external validity of the findings of the multiple case study is the low number of companies and participants from Company A. 
The limited number of people and companies involved means that the results may not be generalizable to other automotive companies or industrial contexts. 
However, the current multiple case study extends and complements a previous work published by the authors, where a multiple case study was structured and run with the same methodology with representatives coming from different companies, widening the scope of the combined results and strengthening their validity. 

Finally, a second possible threat to the external validity of the results of this work is the difference in scope between the automotive field and the other sub-fields of \cps. 
It may be possible that different types of \cps\ may be more ready than vehicles to adopt \CE, but at the best of the authors' knowledge this is not the case. 
Additionally, if this was indeed true, it would be expected that the results of the literature review would have hinted at this possibility. 

\section{Conclusions and Future Work}\label{sec:confw}
\subsection{Conclusions}

This work aimed at formulating an overview of the engagement on the \CE\ practice in the context of \cps, uniting an analysis of the state-of-the-art in research achieved through a systematic literature review to a multiple case study conducted with automotive industrial representatives. 
The resulting impression is a field that has not reached technical maturity yet. 
High-level analysis studies are present in higher numbers than solution proposals and the state-of-practice is yet to be achieved due to the numerous challenges still to be solved. 
However, the prospective gains are definitely appealing for the industrial field. 
It is foreseeable that, as the more abundant conceptual research points at possible solutions to the practical hurdles, in time an increasing number of solutions will be proposed, attempted and validated, thus unlocking the advantages that \CE\ can bring thanks to real-world data-driven software evolution. 

\subsection{Future Work}
As future effort a design study demonstrating a full experimentation cycle is currently in its starting phase. 
The goal is to showcase a prototypical software experimentation procedure conducted on an automotive platform. 
The study is meant to show the feasibility of the approach, starting from the initial software deployment to the systems, to a software variant deployment and execution, data collection, result analysis, and final best-variant adoption.

\section*{Acknowledgements}
The authors wish to thank the colleague Yue Kang for his help during the current multiple case study, the colleague Hang Yin for his support during the previously reported multiple case study, and all the industrial representatives for their time and valuable feedback. 

\section*{Funding}
This work was supported by the projects \textit{COPPLAR Project - CampusShuttle cooperative perception and planning platform}, funded by Vinnova FFI [grant number 2015-04849]; and \textit{Highly Automated Freight Transports}, funded by Vinnova FFI [2016-05413].

\section*{References}
\bibliography{library}

\begin{thebibliography}{16}
\expandafter\ifx\csname natexlab\endcsname\relax\def\natexlab#1{#1}\fi
\providecommand{\url}[1]{\texttt{#1}}
\providecommand{\href}[2]{#2}
\providecommand{\path}[1]{#1}
\providecommand{\DOIprefix}{doi:}
\providecommand{\ArXivprefix}{arXiv:}
\providecommand{\URLprefix}{URL: }
\providecommand{\Pubmedprefix}{pmid:}
\providecommand{\doi}[1]{\href{http://dx.doi.org/#1}{\path{#1}}}
\providecommand{\Pubmed}[1]{\href{pmid:#1}{\path{#1}}}
\providecommand{\bibinfo}[2]{#2}
\ifx\xfnm\relax \def\xfnm[#1]{\unskip,\space#1}\fi
\bibitem[{Auer and Felderer(2018)}]{AF18}
\bibinfo{author}{Auer, F.}, \bibinfo{author}{Felderer, M.},
  \bibinfo{year}{2018}.
\newblock \bibinfo{title}{Current state of research on continuous
  experimentation: A systematic mapping study}, in: \bibinfo{booktitle}{2018
  44th Euromicro Conference on Software Engineering and Advanced Applications
  (SEAA)}, \bibinfo{organization}{IEEE}. pp. \bibinfo{pages}{335--344}.
\newblock \DOIprefix\doi{10.1109/SEAA.2018.00062}.
\bibitem[{Bosch(2012)}]{Bosch12}
\bibinfo{author}{Bosch, J.}, \bibinfo{year}{2012}.
\newblock \bibinfo{title}{Building products as innovation experiment systems},
  in: \bibinfo{booktitle}{International Conference of Software Business},
  \bibinfo{organization}{Springer}. pp. \bibinfo{pages}{27--39}.
\bibitem[{Bosch and Eklund(2012)}]{BE12}
\bibinfo{author}{Bosch, J.}, \bibinfo{author}{Eklund, U.},
  \bibinfo{year}{2012}.
\newblock \bibinfo{title}{Eternal embedded software: Towards innovation
  experiment systems}, in: \bibinfo{booktitle}{International Symposium On
  Leveraging Applications of Formal Methods, Verification and Validation},
  \bibinfo{organization}{Springer}. pp. \bibinfo{pages}{19--31}.
\bibitem[{Eklund and Bosch(2012)}]{EB12}
\bibinfo{author}{Eklund, U.}, \bibinfo{author}{Bosch, J.},
  \bibinfo{year}{2012}.
\newblock \bibinfo{title}{Architecture for large-scale innovation experiment
  systems}, in: \bibinfo{booktitle}{2012 Joint Working IEEE/IFIP Conference on
  Software Architecture and European Conference on Software Architecture},
  \bibinfo{organization}{IEEE}. pp. \bibinfo{pages}{244--248}.
\bibitem[{Fagerholm et~al.(2017)Fagerholm, Guinea, M{\"a}enp{\"a}{\"a} and
  M{\"u}nch}]{FGMM17}
\bibinfo{author}{Fagerholm, F.}, \bibinfo{author}{Guinea, A.S.},
  \bibinfo{author}{M{\"a}enp{\"a}{\"a}, H.}, \bibinfo{author}{M{\"u}nch, J.},
  \bibinfo{year}{2017}.
\newblock \bibinfo{title}{The right model for continuous experimentation}.
\newblock \bibinfo{journal}{Journal of Systems and Software}
  \bibinfo{volume}{123}, \bibinfo{pages}{292--305}.
\bibitem[{Giaimo et~al.(2019)Giaimo, Andrade and Berger}]{GAB19}
\bibinfo{author}{Giaimo, F.}, \bibinfo{author}{Andrade, H.},
  \bibinfo{author}{Berger, C.}, \bibinfo{year}{2019}.
\newblock \bibinfo{title}{The automotive take on continuous experimentation: A
  multiple case study}, in: \bibinfo{booktitle}{2019 45th Euromicro Conference
  on Software Engineering and Advanced Applications (SEAA)},
  \bibinfo{organization}{IEEE}. pp. \bibinfo{pages}{126--130}.
\newblock \DOIprefix\doi{10.1109/SEAA.2019.00028}.
\bibitem[{Giaimo and Berger(2017)}]{GB17}
\bibinfo{author}{Giaimo, F.}, \bibinfo{author}{Berger, C.},
  \bibinfo{year}{2017}.
\newblock \bibinfo{title}{Design criteria to architect continuous
  experimentation for self-driving vehicles}, in: \bibinfo{booktitle}{2017 IEEE
  International Conference on Software Architecture (ICSA)},
  \bibinfo{organization}{IEEE}. pp. \bibinfo{pages}{203--210}.
\bibitem[{Giaimo et~al.(2017)Giaimo, Berger and Kirchner}]{GBK17}
\bibinfo{author}{Giaimo, F.}, \bibinfo{author}{Berger, C.},
  \bibinfo{author}{Kirchner, C.}, \bibinfo{year}{2017}.
\newblock \bibinfo{title}{Considerations about continuous experimentation for
  resource-constrained platforms in self-driving vehicles}, in:
  \bibinfo{booktitle}{European Conference on Software Architecture},
  \bibinfo{organization}{Springer}. pp. \bibinfo{pages}{84--91}.
\bibitem[{Hiller(2016)}]{URL_FUSE_Hiller}
\bibinfo{author}{Hiller, M.}, \bibinfo{year}{2016}.
\newblock \bibinfo{title}{{Thoughts on the Future of the Automotive Electronic
  Architecture}}.
\newblock \URLprefix
  \url{http://www.fuse-project.se/final-seminar-presentation-33558763}.
  \bibinfo{note}{\textit{Accessed 2019-10-22}}.
\bibitem[{Kitchenham et~al.(2015)Kitchenham, Budgen and Brereton}]{KBB15}
\bibinfo{author}{Kitchenham, B.A.}, \bibinfo{author}{Budgen, D.},
  \bibinfo{author}{Brereton, P.}, \bibinfo{year}{2015}.
\newblock \bibinfo{title}{Evidence-Based Software Engineering and Systematic
  Reviews}.
\newblock \bibinfo{publisher}{Chapman \& Hall/CRC}.
\bibitem[{{Lee}(2008)}]{Lee08}
\bibinfo{author}{{Lee}, E.A.}, \bibinfo{year}{2008}.
\newblock \bibinfo{title}{Cyber physical systems: Design challenges}, in:
  \bibinfo{booktitle}{2008 11th IEEE International Symposium on Object and
  Component-Oriented Real-Time Distributed Computing (ISORC)}, pp.
  \bibinfo{pages}{363--369}.
\newblock \DOIprefix\doi{10.1109/ISORC.2008.25}.
\bibitem[{Mattos et~al.(2018)Mattos, Bosch and Olsson}]{MBO18}
\bibinfo{author}{Mattos, D.I.}, \bibinfo{author}{Bosch, J.},
  \bibinfo{author}{Olsson, H.H.}, \bibinfo{year}{2018}.
\newblock \bibinfo{title}{Challenges and strategies for undertaking continuous
  experimentation to embedded systems: Industry and research perspectives}, in:
  \bibinfo{booktitle}{International Conference on Agile Software Development},
  \bibinfo{organization}{Springer}. pp. \bibinfo{pages}{277--292}.
\bibitem[{Olsson and Bosch(2013)}]{OB13}
\bibinfo{author}{Olsson, H.H.}, \bibinfo{author}{Bosch, J.},
  \bibinfo{year}{2013}.
\newblock \bibinfo{title}{Post-deployment data collection in software-intensive
  embedded products}, in: \bibinfo{booktitle}{International Conference of
  Software Business}, \bibinfo{organization}{Springer}. pp.
  \bibinfo{pages}{79--89}.
\bibitem[{Olsson and Bosch(2014)}]{OB14}
\bibinfo{author}{Olsson, H.H.}, \bibinfo{author}{Bosch, J.},
  \bibinfo{year}{2014}.
\newblock \bibinfo{title}{From opinions to data-driven software r\&d: A
  multi-case study on how to close the'open loop'problem}, in:
  \bibinfo{booktitle}{2014 40th EUROMICRO Conference on Software Engineering
  and Advanced Applications}, \bibinfo{organization}{IEEE}. pp.
  \bibinfo{pages}{9--16}.
\bibitem[{Ros and Runeson(2018)}]{RR18}
\bibinfo{author}{Ros, R.}, \bibinfo{author}{Runeson, P.}, \bibinfo{year}{2018}.
\newblock \bibinfo{title}{Continuous experimentation and a/b testing: A mapping
  study}, in: \bibinfo{booktitle}{Proceedings of the 4th International Workshop
  on Rapid Continuous Software Engineering}, \bibinfo{publisher}{ACM},
  \bibinfo{address}{New York, NY, USA}. pp. \bibinfo{pages}{35--41}.
\newblock \DOIprefix\doi{10.1145/3194760.3194766}.
\bibitem[{Runeson and H{\"o}st(2009)}]{RH09}
\bibinfo{author}{Runeson, P.}, \bibinfo{author}{H{\"o}st, M.},
  \bibinfo{year}{2009}.
\newblock \bibinfo{title}{Guidelines for conducting and reporting case study
  research in software engineering}.
\newblock \bibinfo{journal}{Empirical software engineering}
  \bibinfo{volume}{14}, \bibinfo{pages}{131}.

\end{thebibliography}

\end{document}